\documentclass[12pt,preprint]{aastex}

\slugcomment{}

\shorttitle{Globular cluster relative age-dating methods}
\shortauthors{Mar\' \i n-Franch et al.}

\begin{document}

\title{The impact of enhanced He and CNONa abundances on globular cluster relative age-dating methods}

\author{Antonio Mar\' \i n-Franch}
\affil{University of La Laguna and Instituto de Astrof\'\i sica de Canarias, E-38200 La Laguna, Tenerife, Spain
\email{amarin@iac.es} }

\author{Santi Cassisi}
\affil{Osservatorio Astronomico di Teramo, Via M. Maggini, 64100 Teramo, Italy
\email{cassisi@oa-teramo.inaf.it} }

\author{Antonio Aparicio}
\affil{University of La Laguna and Instituto de Astrof\'\i sica de Canarias, E-38200 La Laguna, Tenerife, Spain
\email{antapaj@iac.es} }

\author{Adriano Pietrinferni}
\affil{Osservatorio Astronomico di Teramo, Via M. Maggini, 64100 Teramo, Italy
\email{pietrinferni@oa-teramo.inaf.it} }

\begin{abstract}

The impact that unrecognised differences in the chemical patterns of Galactic 
globular clusters have on their relative age determinations is studied. The two 
most widely used relative age-dating methods, horizontal and vertical, together 
with the more recent relative MS-fitting method, were carefully analyzed on a 
purely theoretical basis. The BaSTI library was adopted to perform the present 
analysis. We find that relative ages derived using the horizontal and  vertical 
methods are largely dependent on the initial He content and heavy element 
distribution. Unrecognized cluster-to-cluster chemical abundance differences 
can lead to an error in the derived relative ages as large as $\sim$0.5 
(or $\sim$6 Gyr if an age of 12.8 Gyr is adopted for normalization), and even larger 
for some extreme cases. It is shown that the relative MS-fitting method is by 
far the age-dating technique for which undetected cluster$-$to$-$cluster 
differences in the He abundance have less impact. Present results are used in order 
to pose  constraints on the maximum possible spread in the He and CNONa 
elements abundances on the basis of the estimates - taken from the literature - 
of the Galactic globular clusters relative age dispersion obtained with the various 
relative age-dating techniques. Finally, it is shown that the age--metallicity 
relation found for  young Galactic globular clusters by the GC Treasury program 
is a real age sequence and cannot be produced by variations in the He and/or heavy 
element distribution.

\end{abstract}


\section{Introduction}

Galactic globular clusters (GGCs) are the most ancient objects known for which 
reliable ages can be determined, and as the Universe cannot be younger than the
 oldest objects it contains,  GGCs provide one of the most robust constraints
 that we have on cosmological models. However, absolute GGCs ages have to be estimated in order 
 to apply this constraint. Although significant improvements  in the absolute GGCs age estimates
 have been obtained in the last decade, they are
 still affected by both observational and theoretical uncertainties \citep{VSB96,
 Cass98, Cass09} at the $\approx20$\% level. Nevertheless, it is possible to determine
 relative GGC ages with the accuracy required to address some outstanding problems, such as 
those related to the Milky Way's formation process. The pivotal importance of these problems 
and the need to improve   age estimates as far  as possible are the basis of the huge
 effort devoted in recent  decades to the gathering of the relative ages of GGCs. As a consequence, 
there exists  quite a rich literature dedicated to this fundamental topic \citep[][and references 
therein]{Stet96, Sar97, R99, SW02, DeA05, MF09}.

The relative age$-$dating techniques for GGCs - almost universally adopted - can be 
separated into two basic classes: those methods that are based on brightness difference 
measurements - the {\sl vertical} method -, and those that are based on color difference 
measurements - the {\sl horizontal} method - in the color--magnitude diagram (CMD). The most 
commonly adopted of the {\sl vertical} methods is the magnitude difference between the main sequence 
turn-off (MSTO) and the zero age horizontal branch (ZAHB), usually estimated starting from the 
level of the RR Lyrae instability strip. The {\sl horizontal} method is based on the 
measurement of the color difference between the MSTO and a point in the lower part of 
the red giant branch (RGB).

It is clear that both approaches are independent of distance, reddening, and uncertainty 
in the adopted photometric zero points. However, apart from this common characteristic, 
the two methods present different advantages and drawbacks. In particular, from the point 
of view of their theoretical calibration, although the dependence of the ZAHB luminosity 
on the metallicity is still controversial, the vertical method seems to be more reliable 
than the horizontal one, which is strongly affected by the non-negligible uncertainties 
related to the treatment of superadiabatic convection and color-$T_{eff}$ relations. As a 
result, relative ages determined using the horizontal method are strongly model-dependent. 
On the other hand, from the observational point of view, the horizontal method seems to be 
largely unaffected - or in any case to quite a minor extent - by the difficulty of measuring 
the MSTO brightness due to its verticality in the CMD. An accurate determination of 
the ZAHB luminosity can also be a thorny problem, in particular for  clusters at the extreme 
boundaries of the metallicity distribution, which generally have a red (or blue) horizontal 
branch and no RR Lyrae stars. In both methods, in order to minimize both observational and 
theoretical uncertainties in the relative age determination, the magnitude or the color 
difference estimated for a GGC is compared with that of a GGC of similar metallicity. A 
detailed discussion of the advantages and disadvantages of all the differential age-dating 
methods can be found in \citet{Stet96} and we refer the interested reader to the quoted reference.

\citet{MF09} have quite recently performed a detailed analysis of the relative ages of a 
sizeable sample of GGCs. In order to increase the level of accuracy in their age estimates 
they have developed an independent method for estimating the MSTO brightness difference 
between two GGCs of similar metallicity. This method is based on a simultaneous fit of the
fainter portion of the MS - whose location is quite insensitive to cluster age but largely 
dependent on cluster metallicity - and the lower portion of the RGB - whose location is 
strongly dependent on the cluster chemical composition (hereinafter, the {\sl relative
 MS-fitting}, rMSF method). A careful analysis of the advantages and uncertainties of 
this method has been performed by \citet{MF09} and  will not be repeated here.

It is important to note that all relative age-dating methods are based on the implicit 
assumptions that all GGCs, apart from the well known differences in their iron content and, 
eventually, in age, are relatively similar objects, i.e., with negligible - if any - 
differences in the helium content and/or heavy elements distribution. However, this common 
idea has been severely challenged in recent decades by the growing evidence, based on accurate 
spectroscopical measurements, that quite peculiar chemical patterns among various GGCs with 
the same iron content do exist, and quite often also among stars belonging to the same cluster 
\citep[for a complete review on this issue we refer the reader to][]{Gratton04}.

Even more surprising has been the recent discovery that many GGCs host multiple stellar 
populations \citep{Bedin04, Piotto05, Piotto07, Milone08, Cass08}, whose photometric 
properties can be understood in terms of sometimes quite large differences in the 
initial He content and/or in the heavy elements distributions and/or age. On the basis 
of the quoted peculiar photometric and spectroscopic properties, a possibility now considered more 
plausible is that the multi-population evidence is not a peculiarity of a 
restricted number of objects, but on the contrary, that it could be a common feature 
in the GGC system. The main difference is that some clusters could have the capability of
retaining the different stellar populations, whereas others could at present be formed 
completely \citep{Derc08, DanCal08}, or at a significant level (between 50 and 70\%), by 
second generation stars \citep{Carret09}.

It is evident that the presence of observationally deceptive differences in the chemistry 
of GGCs might have a huge impact on relative age measurements. In fact, observational 
differences between two GGCs of similar metallicity, as detected with any of the quoted 
techniques, could be erroneously interpreted as due to an age difference when - on the 
contrary - it might only be due to an unrecognized difference in the chemical abundance pattern. 

This issue has been recently investigated by \citet{dantona09}, who have only considered 
the specific case of the two GGCs (NGC~1851 and NGC~6121) and showed that the brightness 
difference existing between the sub-giant branches of these two GGCs can be explained 
as due to a difference in the CNO element abundance, even under the hypothesis that the
 two clusters are perfectly coeval.

In this article, we wish to address this issue in greater detail and more extensively, but 
only on purely theoretical grounds. It is worth noting that we do not investigate the 
robustness of the theoretical calibration of each independent age-dating method. We 
use a homogeneous set of stellar models in a completely differential approach for 
estimating how unrecognized differences in the chemical patterns of GGCs affect the 
relative age results obtained with the commonly adopted methods.

The plan of this paper is as follows: in the next section, the adopted theoretical 
framework is presented; in Section~\ref{methodsSec} we briefly review the relative 
age$-$dating techniques that we'll take into account in present work; in Section~\ref{results}
we present our analysis by discussing individually each age-dating method; and we close with our conclusions 
and final remarks.

\section{The theoretical framework}\label{isochrones}

The present analysis has been performed by using the BaSTI\footnote{The whole library of 
stellar evolutionary predictions adopted in this work, as well as additional stellar 
model predictions can be retrieved at the following URL address: 
http://www.oa-teramo.inaf.it/BASTI.} library of stellar models and isochrones computed 
for the $\alpha$-enhanced mixture presented by \citet{Pie06}. These models cover the 
whole metallicity range of GGCs, and have been computed by assuming a primordial He 
content equal to Y=0.245 \citep[see][and references therein]{Cass03} and a He 
enrichment ratio $\Delta{Y}/\Delta{Z}=1.4$. From now on, these will be referred to as
 {\sl reference} isochrones. For a detailed discussion of the adopted physical 
inputs we refer to \citet{Pie04, Pie06}, while for a careful discussion of the adopted 
color-$T_{eff}$ relation and bolometric scale we refer to \citet{Cass04} and
\citet{bedin05}.

For the aim of testing the impact of an enhanced He content, we have computed an 
additional extended set of stellar models for low-mass stars for both the H- and 
He-burning stages, by adopting the same $\alpha$-enhanced mixture but accounting for 
three larger He contents; namely, Y=0.30, 0.35, and 0.40. 

In this context,  it is worth mentioning that for a fixed global metallicity (Z), a 
change in the adopted He content causes a variation in the corresponding iron content, 
[Fe/H], according to the well-known definition of [Fe/H] as a function of Z and Y. 
Usually this change is very small because the He contents adopted in stellar model 
computations do not differ greatly from the \lq{canonical}\rq\ value ($0.245\le{Y}\le0.27$). 
However, when dealing with extremely He rich populations, this effect might not be 
completely negligible. Therefore, when computing models for a fixed [Fe/H] and a 
given He abundance, we rescale the global metallicity Z in order to preserve the 
[Fe/H] value: the metallicity Z has to be reduced when the He content is increased. 
This choice allows us, when comparing isochrones for various He-content assumptions, 
to compare isochrones for the same iron content consistently.

We recognize that the adopted He enhancements with respect to the canonical value 
($Y\sim0.25$) could appear too large to be considered realistic for GGCs. 
However, we emphasize that, in the present analysis, we want to investigate - on a purely
theoretical basis - the dependence of the relative age determinations on unrecognized 
peculiar chemical patterns. For this reason, we also take into account these extreme He enhancements.

Finally, for the purpose of checking the impact of a \lq{peculiar\rq\ heavy 
element distribution on relative age-dating methods, we have also adopted 
the set of stellar models presented in \citet{Pie09} accounting for an extreme 
CNONa chemical patterns. In order to cover the whole metallicity range
 sampled by GGCs properly, we have extended the original set of models to lower
metallicities, so the final database of stellar models adopted in the present 
analysis covers an iron content range from $-2.89$ to $-0.56$.

For a detailed description of the assumed heavy element distribution and more 
details on these stellar models and isochrones we refer the reader to the quoted reference.
 However, it is important to remark that the adopted heavy element distribution
 corresponds to a mixture in which the sum (C$+$N$+$O) is enhanced by a factor
 of approximately 2 with respect to the {\sl reference} mixture. This value
is consistent - although it represents an upper limit - with the results of the 
spectroscopic analysis performed by \citet{Carret05} for the extreme values of 
the chemical anti-correlations observed in GGCs.

It is worth noting that the stellar models for the various chemical compositions 
adopted in the present work are based on the most up-to-date physics currently available 
and, more importantly for our aim, all evolutionary predictions are fully 
homogeneous and self-consistent, being based exactly on the same physical framework.

Before closing this section, we wish to note that it is quite important to 
perform the comparison between {\sl reference} isochrones and those corresponding 
to the \lq{peculiar}\rq\ chemical patterns at fixed iron content. The main 
reasons for this choice are the following: {\sl i)} when performing relative GGC 
age measurements the commonly adopted procedure is to divide the whole cluster
sample into sub-samples on the basis of their [Fe/H] value (that is, the parameter 
provided by spectroscopical measurements) and then to apply the adopted relative 
age dating method to each selected sub-sample; {\sl ii)} at fixed global metallicity 
Z, and hence [M/H], a change in the helium content (see above) and/or in the heavy
 element distribution \citep[see the discussion in ][]{Pie09} implies a change in
the corresponding [Fe/H] value. Therefore, in order to obtain reliable results, we need 
to use the same approach that is adopted when managing real stellar systems; 
that means investigating the impact of changing the helium content or the heavy 
element distribution at fixed [Fe/H].

\section{Globular cluster relative age-dating methods}\label{methodsSec}

In this section, the most widely used relative age-dating methods are described. 
Figure~\ref{methods} illustrates the horizontal, vertical and rMSF techniques, 
applied to 8, 10, 12, and 14 Gyr {\sl reference} isochrones with the same metallicity. $Hubble$ 
$Space$ $Telescope$ ACS/WFC filters $F606W$ ($\sim V$) and $F814W$ ($\sim I$) are used here as a guideline. 

\subsection{Horizontal method}\label{Hmethod}

\citet{VBS90} were pioneers in measuring GGC relative ages by making use of the horizontal 
method. They essentially derived relative ages comparing the color difference between the 
MSTO and the location of a well defined point in the lower RGB, located 2.5 magnitudes brighter 
than a point in the upper main sequence (MS) that is 0.05 magnitudes redder than the MSTO. 
From now on this color difference will be referred to as the horizontal parameter. This method is 
illustrated in the left hand panel of Figure~\ref{methods}, where the dependence of the horizontal 
parameter on age is clearly shown. 

Using the {\sl reference} isochrones described in Section~\ref{isochrones}, the horizontal 
parameter was computed for a wide range of ages (from 6 to 15 Gyr) and iron content (from 
about $-2.6$ to $-0.3$). The results are plotted in the upper panel of Figure~\ref{grids}. 
It shows the resulting horizontal parameter as a function of [Fe/H]. Lines represent model 
horizontal parameter in steps of 1 Gyr (solid lines) and 0.5 Gyr (dotted lines). 
This theoretical grid will be used to derive relative ages based on the horizontal method. 
To do this, the curves are interpolated using a spline surface, and therefore, ages can be 
retrieved by a direct comparison of the target isochrone's (or GGC's) horizontal parameter 
with the interpolated surface. 

To evaluate the impact that unrecognized differences in the chemical patterns of 
GGCs have on the relative age determination, one can measure the horizontal parameter for 
an isochrone of given age but with a different He content and/or heavy element distribution 
(with respect to the {\sl reference} set). By doing so, the age corresponding to that measured
 horizontal parameter can be retrieved. Finally, comparing the input with the retrieved isochrone age, 
it is possible to evaluate the impact of {\sl undetected} chemical abundances
   differences on the age determined with the horizontal method.

\subsection{Vertical method}\label{Vmethod}

Potentially, more reliable age indicators are those related to the brightness of the MSTO, 
in particular the vertical method. The first surveys of GGC relative ages using the vertical
 method were carried out by \citet{G87, P87} and \citet{SK89}, who used the brightness difference
  between the ZAHB and the MSTO (the vertical parameter from now on) to estimate the ages of sizeable 
  samples of GGCs. This method is illustrated in the central panel of Figure~\ref{methods}, where 
  the vertical parameter dependence on age is shown. Since  it is well known the ZAHB brightness 
  level for old stellar systems, such as GGCs, is independent of age, its level is the same for the 
  four plotted {\sl reference} isochrones.

Following the same methodology as for the horizontal method, {\sl reference} isochrones were used 
to determine the vertical parameter as a function of age and [Fe/H]. The results are plotted in the 
central panel of Figure~\ref{grids}. Lines represent the resulting vertical parameter in steps of 
1 Gyr (solid lines) and 0.5 Gyr (dotted lines). Again, the curves are interpolated using a spline 
surface and will be used to derive relative ages based on the vertical method.  Once the 
theoretical grid is set up, the age of a GGC - or of an isochrone with a different He content and/or 
heavy elements distribution - can be determined from its vertical parameter by comparing it with the
 {\sl reference} isochrone grid. 

\subsection{Relative MS-fitting method}\label{rMSFmethod}

The rMSF method was first used to derive relative ages for a large, homogeneous database of GGC 
photometry by \citet{MF09}. The method is discussed in detail in the quoted reference and only a 
brief description is presented here. The CMD location of the faint MS of a GGC is independent of its 
age, but highly dependent on its metallicity. For this reason, MS and RGB fitting between clusters 
with similar metallicity is performed. The right hand panel of Figure~\ref{methods} shows an example  
rMSF in which the 10, 12, and 14 Gyr isochrones have been shifted in both magnitude and color to fit 
the CMD location of the 8 Gyr one, all isochrones having the same metallicity. The fit is performed 
following the prescriptions of \citet{MF09}; that is, in a least-squares fashion and taking into account two 
CMD regions that have little dependence on cluster age. These regions are shaded in Figure~\ref{methods}. 
It can be seen how the rMSF method provides gives the relative brightness of the considered isochrones' MSTOs. 

Again, the {\sl reference} isochrones described in Section~\ref{isochrones} were used to compute the 
theoretical grid. The lower panel of Figure~\ref{grids} shows the model MS turn-off in the F606W filter 
$M_{F606W}^{\rm MSTO}$ as a function of [Fe/H]. Lines represent MSTO magnitudes in steps of 1 Gyr (solid 
lines) and 0.5 Gyr (dotted lines). The curves are interpolated using a spline surface so that one can easily 
estimate $M_{F606W}^{\rm MSTO} = f({\rm [Fe/H], age})$. Finally, relative ages can be estimated by using the 
interpolated surface.

The rMSF method was be applied to our set of isochrones, with different a He content and/or heavy element 
distribution, by shifting in each case the considered isochrone in both color and magnitude to fit the 
corresponding same metallicity {\sl reference} one. This procedure provides the brightness of the MSTO 
relative to the {\sl reference} isochrone's MSTO, which can be used to derive the relative age as previously described.
 
\section{Results}\label{results}

The impact of an enhanced initial He content or an extreme CNONa mixture on the relative ages obtained by the
 different dating methods is illustrated in Figure~\ref{HeCNOEffect}. Horizontal, vertical, and rMSF methods 
 applied to isochrones with the same [Fe/H], same age, but different values of Y (upper panels) and different 
 CNONa abundances (lower panel) are shown. The upper left hand panel shows how, for the same age and same 
 [Fe/H], the horizontal parameter increases when decreasing Y.\footnote{We wish to note that this monotonic 
 behavior of the horizontal parameter as a function of the He content is not present in the whole explored 
 metallicity and/or age range. On the basis of our own computations we have verified that this occurrence
  is related to the different dependence of the effective temperature of the MS TO and of the RGB on the He 
  content. We have verified that this evidence is also supported by other independent evolutionary computations 
  such as those by \citet{berte08} \citep[see also ][]{catel09}.} This means that the horizontal-method 
  derived age does not coincide with the isochrone (input) one. It depends on the Y value, the derived 
  age being older for larger values of Y. The upper central panel illustrates the effect of Y on the vertical method. 
  The large impact that the He content has on the ZAHB level, as well as on the MSTO brightness,  can be seen. 
  This translates into a huge effect on the vertical parameter and consequently on the derived ages. It is
   worth mentioning that the ZAHB and MSTO magnitude shifts as a consequence of Y variation do not compensate 
   each other. On the contrary, increasing Y translates into a fainter MSTO and a brighter ZAHB. The effect of 
   varying the He abundance on the rMSF method is illustrated in the upper right hand panel. Increasing Y
    makes the MSTO fainter, as well as the MS locus. According to the figure, both effects compensate each
     other during the rMSF procedure, so the method turns out to be rather insensitive to He variations.

Similar arguments can be made for the CNONa variations based on the lower panels. The lower left hand panel 
illustrates the effect that a different CNONa abundance has on the horizontal-method derived age. In the example
 of the figure, the horizontal parameter increases if an extreme CNONa mixture is considered, and this translates 
 into a retrieved age that is younger than the input one. The lower central panel shows how the vertical parameter
 increases in the case of an extreme CNONa mixture. This translates into older vertical-method derived ages. 

We wish to note that the ZAHB brightness level is also affected by the CNONa elements enhancement, becoming 
slightly brighter - at least in the case of the adopted CNONa enhancement - at fixed iron content \citep[see 
the discussion in ][]{Pie09}. Needless to say, that this effect is properly taken into account when investigating
 the effect of a change in the CNONa distribution on the vertical-method. However, for the sake of clarity this 
 effect is not shown in Figure~\ref{HeCNOEffect}, since for the selected iron content the ZAHB brightness difference
  is only of $\sim0.01$ mag.

Finally, the effect of varying the heavy element distribution on the rMSF method is shown in the lower right 
hand panel. It can be seen that the CNONa isochrones shows a relative MSTO slightly fainter (an occurrence that
 mimics an older age if one ignores the \lq{true}\rq\ age and assumes the same heavy element mixture) as the 
 reference isochrone. 

\subsection{Horizontal and Vertical methods}

\subsubsection{Helium}

For a more general analysis, the age of a set of isochrones with different values of Y was determined 
using the horizontal and vertical methods. A wide range of ages (8, 10, 12, and 14 Gyr) and total metallicities 
([Fe/H]=$-$2.62, $-$2.14, $-$1.62, $-$1.31, $-$1.01, and $-$0.70) was considered. The results are shown in 
Figure~\ref{classicalHe}. The left hand panels show the horizontal-method derived relative age as a function of 
[Fe/H], for different Y values. An age of 12.8 Gyr has been adopted for normalization \citep{MF09}. It can be 
seen how the obtained age coincides with the isochrone (input) age for Y$\sim$0.25 (black line) as expected 
because these are the {\sl reference} isochrones used to build the theoretical grids shown in Figure~\ref{grids}. 
This is not the situation if larger values of Y are considered. It is clear that even a value of Y$\sim$0.30 
(significantly larger than the canonically adopted value, but not completely ruled out - at least for the 
sub-populations hosted by some GGCs according to the recent observational evidence of GGCs multi-populations) 
translates in derived horizontal-method relative ages that are from 0.08 to 0.3 older than the input ones (or 
1 to 4 Gyr, if these values are transformed to absolute ages), for iron contents larger than [Fe/H]$\ge$ $-$1.01. 
This age discrepancy is clearly larger for larger values of Y. For [Fe/H]$\le$ $-1.31$, the dependence of the 
horizontal method results in the He content seeming to be significantly lower. \cite{VBS90} have already analyzed the 
dependence of the horizontal method on Y and [O/Fe] and reached similar conclusions. They noted that the 
horizontal method becomes quite uncertain for metallicities larger than [M/H]$\sim$ $-$1.2, or  iron content 
[Fe/H]$\sim$ $-$1.6. 

The right hand panels of Figure~\ref{classicalHe} show the vertical-method derived age as a function of [Fe/H] 
for different Y values. These ages have been determined for the same set of isochrones as in the previous section 
but, for clarity, only 8 and 10 Gyr input ages are plotted. In this case, the obtained relative ages are from 0.2 
to 0.4 (or 3 to 5 Gyr in absolute values) older than the isochrone (input) ones if Y=0.30 is considered 
independently of the iron content. This result worsens if larger values of Y are taken into account, reaching 
relative age differences of the order of more than 0.8 (or 10 Gyr) if extreme Y=0.35 or 0.40 values are considered, 
as can be seen in the figure. 

\subsubsection{Extreme CNONa mixture}

A similar analysis was done to evaluate the impact of extreme CNONa  mixtures. Figu\-re~\ref{classicalCNO} 
shows the results of this analysis. The left hand panels show the horizontal-me\-thod derived relative age as a 
function of [Fe/H] for the reference (canonical) and extreme CNONa mixtures. Horizontal-method relative ages 
obtained in the case of extreme CNONa mixtures tend to be $\sim$ 0.1 (that is $\sim$ 1 Gyr in absolute age) older 
than the input ages, with some exceptions at intermediate iron content. In any case, no clear trend with iron
 content is found. These results differ from the analysis of the horizontal method dependence on [O/Fe] carried 
 out by \cite{VBS90}. They argued that the horizontal method becomes quite uncertain for intermediate- and 
high-metallicity GGCs. In the present analysis it is shown how this method is quite sensitive to the CNONa mixture, 
 independently of the iron content. The origin of this difference is probably due to the differences in the
  adopted heavy element mixtures, as well as in the physical framework (radiative opacity, equation of state, etc.).

The right hand panels shows the vertical-method derived ages for canonical and extreme CNONa isochrones. In this 
case, vertical-method relative ages derived for the extreme CNONa isochrones are significantly older than the 
input ones for [Fe/H]$\ge$ $\sim-2$, the difference increasing with metallicity.  In the high metallicity regime, 
retrieved relative ages are up to $\sim$ 0.3 (or $\sim$ 4 Gyr) older than the input value. 
   
In summary, the horizontal- and especially vertical-method derived ages are largely dependent on the 
initial He value and CNONa mixture. In other words, if the relative age of a GGC is measured using these methods, 
undetected differences in the He content and/or CNONa abundance translates in an unreal age determination. 
The difference between the measured relative age and the actual one can be of the order of $\sim$ 0.4-0.8 
(which correspond to several Gyr in absolute age), especially for high metallicity clusters. It is clear that 
the vertical method more sensitive to variations in Y and/or CNONa chemical abundances. We note that, while
 for the He content a wide interval has been explored, for the case of the extreme CNONa mixture, we have 
 investigated only the case of an enhancement factor equal to 2 in the CNO elements abundance. It is evident 
 that, in case of larger CNO enhancement factors, the expected difference between the input age and the 
 retrieved one would be quite larger.

\subsection{Relative MS-fitting method}

The rMSF-method ages were also measured for our set of isochrones, grouping them in subsets with the 
same iron content. In this case, a larger age interval was considered. Results are shown in 
Figure~\ref{msFittingHe}. Error bars represent the relative age uncertainty derived from the rMSF 
procedure \citep[$\sigma_{MSF}$, described in ][]{MF09}. If He enhanced isochrones are considered, 
it is apparent that the measured rMSF-method relative ages tend to be slightly older than the input 
ones, especially for high metallicity isochrones. If all values of Y and input ages are considered, the 
mean determined relative age is $\sim$0.03 (or 0.4 Gyr if expressed in absolute age) older than the 
input one, showing an rms dispersion around this mean of also $\sim$ 0.03 (or 0.4 Gyr). It is noticeable 
that this result is also independent of the Y value, that is, even extreme Y values translate into a 
$\sim$0.03$\pm$0.03 relative age difference between the input and rMSF-method derived ages. 

In the case of the extreme CNONa mixture, the results are shown in Figure~\ref{msFittingCNO}. In this case, 
it appears that  rMSF-method derived relative ages are $\sim$0.10$\pm$0.02 (or $\sim$1$\pm$0.3 Gyr in 
terms of absolute age) older than the actual ages. No significant trend with [Fe/H] is found.

\section{Discussion}

In order to present the results in a clearer and more concise way, the partial derivatives of the 
relative age with Y and CNONa, with different iron contents, have been calculated. The partial derivatives 
$\delta Age_{NORM} / \delta Y$ have been computed using the mass change in He, that is 
Y=0.25, 0.30, 0.35 and 0.40. For this purpose, and for each value of input age and iron content, a least 
squares fit has been performed to derive the slope of the $Age_{NORM}$ as a function of Y. Finally, results 
for different input ages have been averaged to derive the final $\delta Age_{NORM} / \delta Y$ as a function 
of $[Fe/H]$. Concerning the CNONa, a similar procedure has been followed, but in this case, since the 
CNONa extreme mixture has a sum (C+N+O+Na) that is a factor of 2 (about 0.3 dex) larger than in the reference 
case,  $\delta Age_{NORM} / \delta CNONa$ has been computed by accounting for a $\delta CNONa$ = 0.3.  

The results are shown in Figure~\ref{deriva}, where the partial derivatives of the relative age with Y 
(left panel) and with CNONa (right panel) are plotted for the horizontal (red squares), vertical (blue circles) 
and rMSF (black triangles) methods. The left panel clearly shows that the vertical method is strongly 
dependent on Y for all metallicities, while the horizontal method is very sensitive to Y only for 
metallicities higher that [Fe/H]$>-1.3$. The rMSF method, on the other hand, is quite insensitive to Y 
for the whole metallicity interval except, maybe, for very high iron content. The right panel shows that the 
three methods are not very sensitive to CNONa,  rMSF method having the relative advantage of showing a CNONa 
dependence that is metallicity independent.

In conclusion, it is evident that both the horizontal and the vertical methods are significantly affected by 
undetected differences in both the initial He content and/or heavy element distribution between the stellar 
systems under scrutiny. Between the two methods, the worst one in this context is clearly the vertical method. 
With regard to the rMSF relative age technique, this is almost unaffected by any variation in the helium 
content, and affected by a change in the metal distribution (in particular in the CNO element abundance) as
 the same level of the horizontal and vertical method. On the basis of present analysis, and taking into account the 
 advantages discussed by \citet{MF09}, we consider the rMSF method much more suitable than other techniques 
 for retrieving relative GGCs ages.

\subsection{Delimiting the He dispersion in GGCs}

It has been shown that the horizontal and  particularly the vertical methods provide ages that are very 
sensitive to differences in chemical abundances. That is, if the relative age of a GGC is measured  using 
these methods, undetected differences in the He content and/or CNONa abundance translates into a wrong relative
 age determination. The difference between the measured relative and actual ages  can be of the order
  of $\sim$ 0.4-0.8 (which correspond to several Gyr in absolute age), especially for high-metallicity clusters. 
  It is clear that the vertical method is most sensitive to variations in Y and/or CNONa chemical abundances. 
  This implies that these two relative age-dating methods are not optimum in the case of having GGCs showing 
  different chemical patterns, as has been found observationally. Peculiar chemical patterns have been detected 
  among various GGCs with the same iron content based on accurate spectroscopic measurements \citep[see ][for 
  a thorough review]{Gratton04}. However, this drawback can be used to limit the possible cluster-to-cluster 
  differences in He according to their measured relative ages, and in particular on the basis of the relative age 
  dispersion.

A recent relative age study based on the vertical method has been carried out by \citet{DeA05}, who found a 
relative age dispersion of $\sim$0.05, $\sim$0.08 and $\sim$0.03 (rms) for the low-, intermediate- 
($-1.7<$ [Fe/H] $<-0.8$) and high-metallicity groups of GGCs, respectively. These dispersions could be due 
to a real age dispersion, to cluster-to-cluster He differences, or a combination of both. In any case, 
these values can be used to limit the He dispersion in GGCs to $\sim$0.010, $\sim$0.012 and $\sim$0.004, 
according to Figure~\ref{classicalHe}.
 
\subsection{Can the He and/or CNONa spread account for the observed rMSF-derived relative ages?}

\citet{MF09} performed a relative age study over 64 GGCs based on the rMSF method. They found that the GGC sample 
can be divided into two groups: (i) a population of old, coeval GGCs with an intrinsic relative age dispersion of
 0.03 and no age--metallicity relation, and (ii), a group of younger GGCs with a well defined age--metallicity 
 relation and an intrinsic relative age dispersion with respect to this relation of also 0.03. In this section, 
 the possibility of these age dispersions and age--metallicity relation being caused by cluster-to-cluster 
 He and/or CNONa variations is analysed.

Regarding the He, the main finding of the present paper is that the rMSF method is the relative age-dating 
technique for which undetected differences in the He content has less impact. The impact is small but not negligible. 
It has been shown that undetected differences in the He content translate into a relative age determination 
that is on average, and considering also extreme Y values, $\sim$0.03$\pm$0.03 older than the actual one. But 
in the previous section, the possibility of having a significant He dispersion in GGCs has been ruled out: based 
on vertical-method relative ages results, the cluster-to-cluster He dispersion has been limited to 
$\sim$0.01. This translates into a source of uncertainty at the level of $\sim$0.004 in the derived rMSF-method 
relative ages, which is not significant. That is, neither the young group of GGCs' age--metallicity relation nor 
the 0.03 intrinsic relative age dispersion obtained by \citet{MF09} is produced by a He dispersion in GGCs.

As far as it concerns the existence of a real spread in the CNONa elements in the GGC system, the empirical 
findings, available so far, seem to show that a CNONa enhancement of a factor of 2 (the same adopted in present 
analysis) with respect the standard $\alpha$-enhanced mixture cannot be ruled out \citep{Carret05}. However, 
this evidence is based on few GGCs, and in any case the quoted CNONa enhancement appears - so far - an upper 
limit (Carretta 2009, private communication). 

Bearing in mind this empirical evidence, we can try to use present results in order to put further 
constraints on the existence of a possible spread in the CNONa element enhancement. As discussed in the previous 
section (see Figure~\ref{msFittingCNO}), in the case of an enhancement of a factor of 2 (i.e., 0.3 dex), the 
rMSF-method derived relative ages are $\sim0.10\pm0.02$ older than the actual ones. This value appear 
larger than the 0.03 relative age intrinsic dispersion found by \citet{MF09}: this means that if the retrieved
 age dispersion is only due to a CNONa spread we can limit the \lq{maximum}\rq\ CNONa enhancement to a factor 
 $\sim1.2$ (i.e., $\sim0.18$~dex or $\sim$20\%). Interestingly enough, this estimate is in very good agreement with 
 currently available spectroscopic measurements for GGCs. In other words, the relative age dispersion found by
  \citet{MF09} could be entirely produced by a cluster-to-cluster CNONa dispersion of 20\% or 0.18 dex.

Finally, as a CNONa enhancement of a factor of 2 translates into a relative age change of $\sim$0.1, it cannot
 be the cause of the young group's age--metallicity relation (in which relative ages go from $\sim$1 to $\sim$0.5). 
 As a conclusion, we reassert that the age--metallicity relation found in \cite{MF09} is a real age sequence. 

\section{Conclusions}

In the present work, we have studied the impact that unrecognized differences in the chemical patterns of GGCs have on their 
relative age determinations. The two most widely used relative age-dating methods, horizontal 
and vertical, together with the more recent relative MS-fitting method described in \citet{MF09}, were 
carefully analyzed on a purely theoretical basis. The BaSTI library of stellar models was adopted to perform 
the present analysis, supplemented by additional evolutionary computations for more extreme assumptions about
 the initial He content. Our main conclusions are summarized here:

\begin{itemize}

\item We find that relative ages derived using the horizontal-method are largely dependent on the initial Y value 
and CNONa mixture. Undetected differences in the He content and/or CNONa abundance translates in an unreal age 
determination. The difference between the measured relative age and the actual one is in the range from 0.08 to
 0.3 (or 1 to 4 Gyr, if these values are transformed to absolute ages), this result worsens for high metallicity clusters.

\item For the vertical method, we find that the obtained relative ages are from 0.2 to 0.4 (or from 3 to 5 Gyr in 
absolute values) older than the isochrone (input) ones if a helium enhancement of Y=0.3 is considered,
 independently of the metallicity. This result worsens if larger - more extreme - values of Y are taken into
  account, reaching age differences of the order of more than 0.8 ($\sim$ 10 Gyr) if extreme Y=0.35 or 0.4 values
   are considered.  The vertical method is most sensitive to cluster-to-cluster undetected variations in Y. 

\item We find that the vertical-method can be used to limit the possible cluster-to-cluster differences in 
He according to their measured relative ages. In particular, the Y dispersion in GGCs has been limited to $\sim$0.01.

\item We find that the rMSF method is the relative age-dating technique for which undetected differences in the 
He content has less impact. It has been shown that neither the young group of GGCs' age--metallicity relation nor 
the 0.03 intrinsic relative age dispersion found by \citet{MF09} is produced by a He dispersion in GGCs.

\item When considering the possibility of undetected differences in the CNONa mixture, our results allow us to
 constrain the maximum possible enhancement of CNONa elements, which should be of the order of 1.2; i.e., 
 $\sim0.18$~dex, with respect a \lq{standard}\rq\ $\alpha-$enhanced mixture.

\item The relative age dispersion found by \citet{MF09} could be entirely produced by a cluster-to-cluster
 CNONa dispersion.

\item We reassert that the age--metallicity relation found in \cite{MF09} is a real age sequence.

\item When taking also into account the advantages of the rMSF technique with respect to the other relative age 
dating methods, it appears that, so far, that the rMSF method is the approach that is much to be preferred for 
retrieving GGC chronology.

\end{itemize}

\begin{acknowledgements}

The authors would like to thank the anonymous referee for very constructive comments.  AMF has been supported
 by the Education and Research Ministry of Spain's Juan de la Cierva postdoctoral position. This work has 
 been financially supported by the Instituto de Astrof\' \i sica de Canarias (grant P3-94) and the Education 
 and Research Ministry of Spain (grant PNAYA2004-06343 and Consolider-Ingenio 2010 Program CSD 2006-00070).
  S.C. acknowledges the partial financial support of INAF through the PRIN 2007 grant n. 
  CRA 1.06.10.04: \lq{The local route to galaxy formation}\rq, and of Ministero della Ricerca 
  Scientifica e dell'Universit\'a (PRIN-MIUR 2007).

\end{acknowledgements}

\clearpage

\begin{figure}
\epsscale{1.00}
\plotone{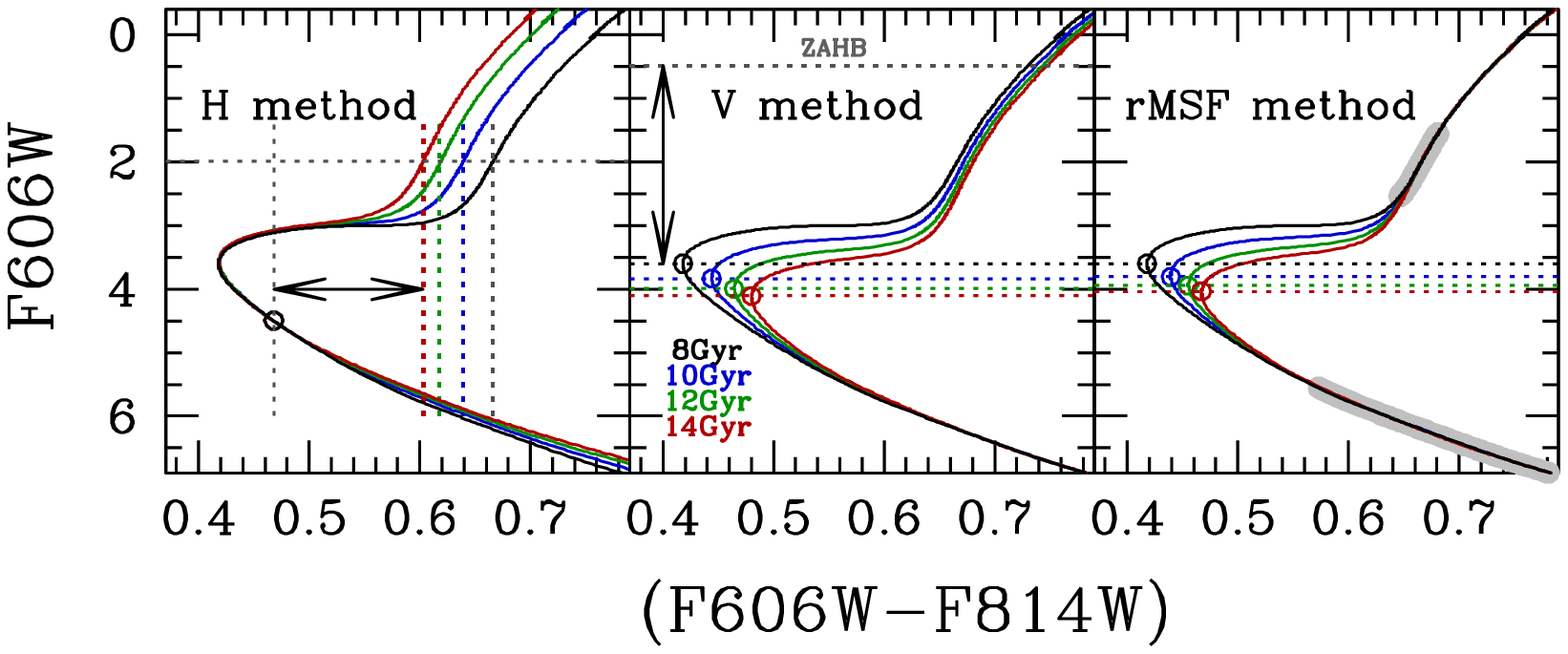}
\caption{he relative dating methods. The BaSTI $\alpha$-enhanced isochrones with 
[Fe/H]=$-1.31$, Y=0.248 and  ages 8, 10, 12 and 14 Gyr are shown. \label{methods}}
\end{figure}

\begin{figure}
\epsscale{1.00}
\plotone{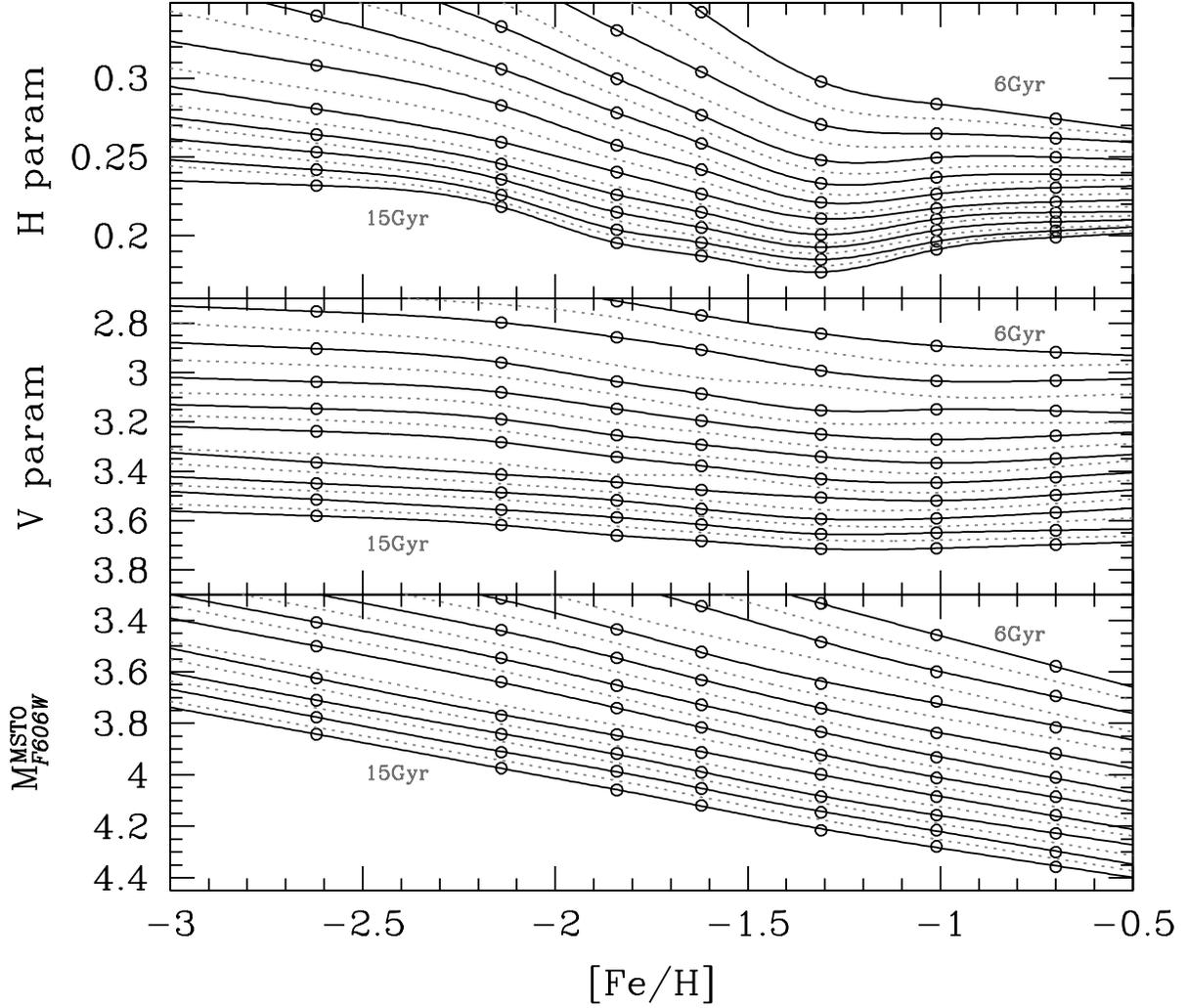}
\caption{Theoretical grids for the H and V parameters, together with the $M_{F606W}^{\rm MSTO}$. 
Open circles represent some of the isochrone measurements used for the grid computation.The adopted 
isochones correspond to the {\sl reference} theoretical grid (see text for more details). \label{grids}}
\end{figure}

\begin{figure}
\epsscale{1.00}
\plotone{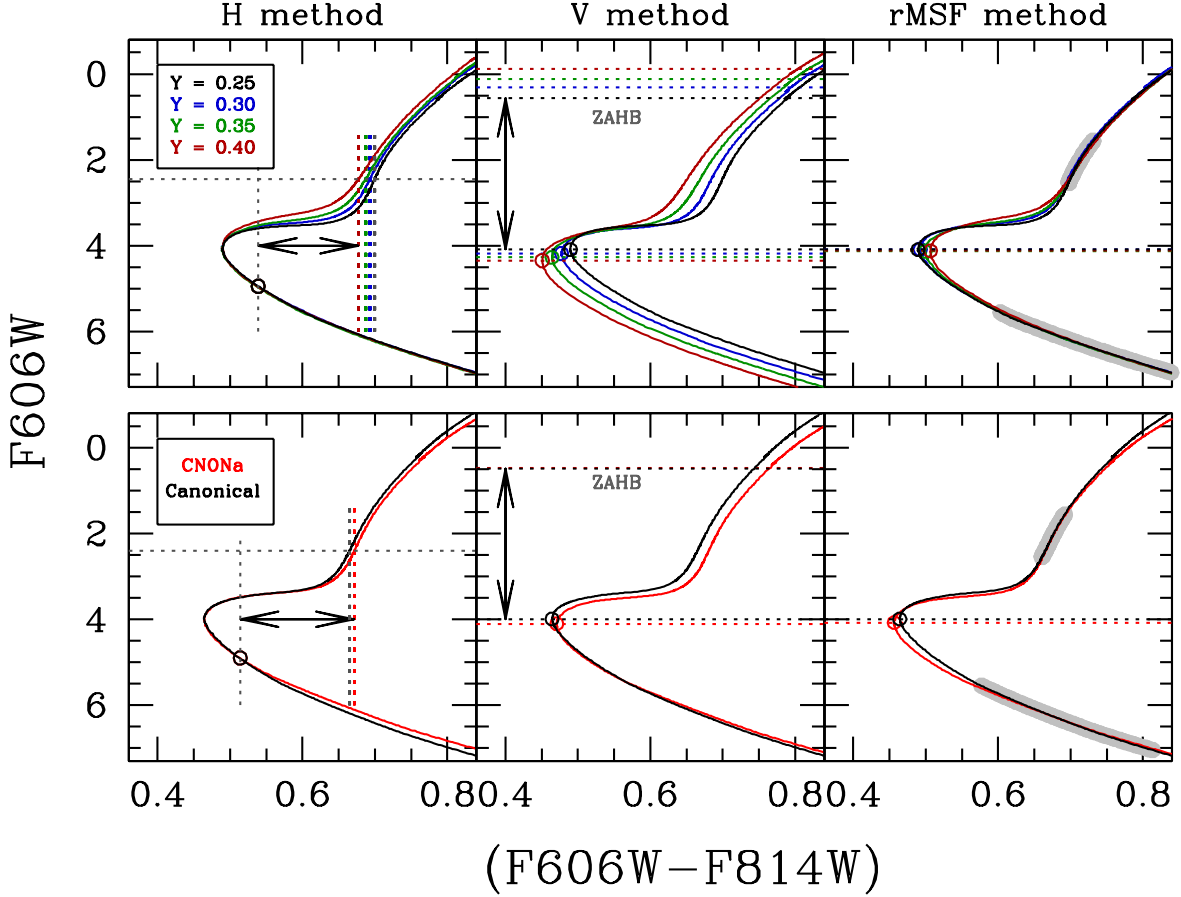}
\caption{Qualitative effect of He (upper panels) and CNONa heavy element (lower panels) abundances 
on the derived ages for the different methods. These are BaSTI $\alpha$-enhanced isochrones with [Fe/H]=$-1.01$ 
(upper panels) and $\sim$ $-1.3$ (lower panels) for an age of 12 Gyr. Different Y values 
(0.25, 0.30, 0.35 and 0.40) and CNONa mixtures (canonical and extreme) are considered. The set 
of colors adopted to mark the different helium and CNONa abundances have been labeled in the figure 
(boxes), and is the same set of colors adopted in the following figures.  \label{HeCNOEffect}}
\end{figure}

\begin{figure}
\epsscale{1.00}
\plotone{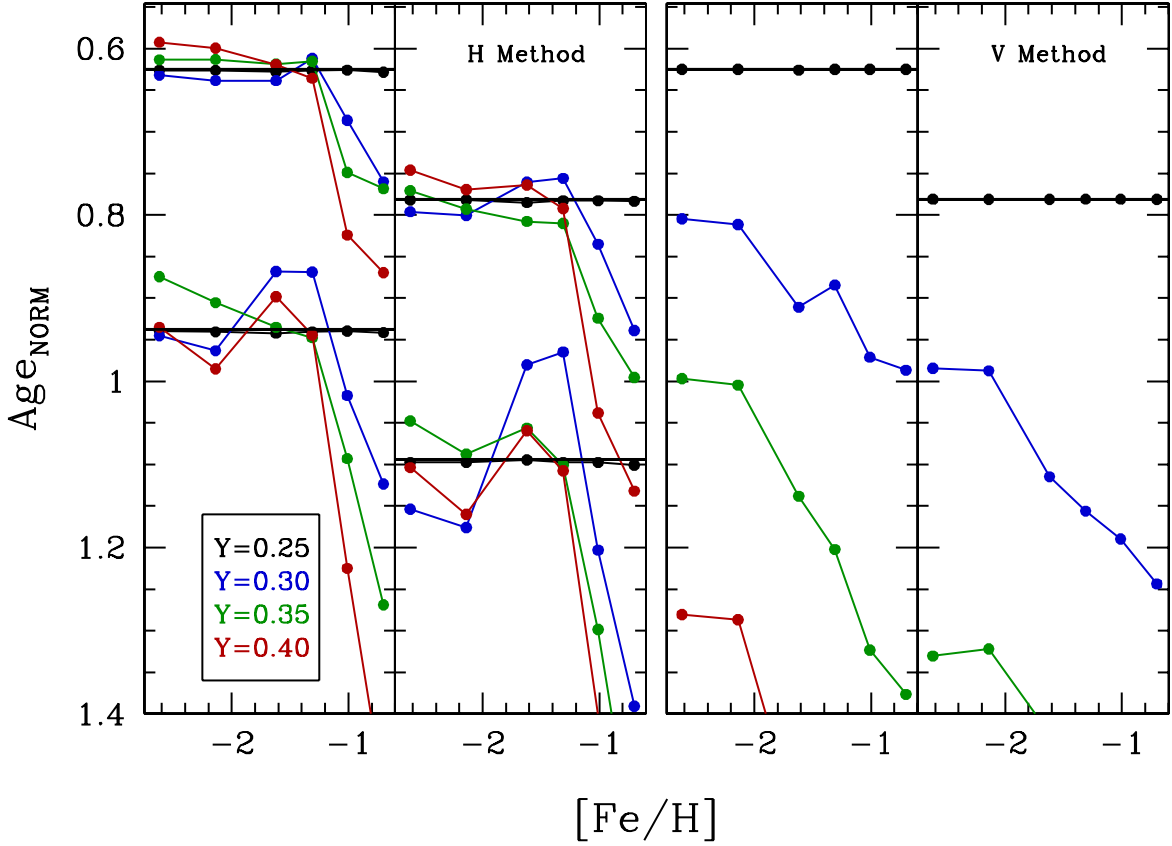}
\caption{Retrieved relative ages measured using the horizontal (left hand panels) and the vertical 
(right hand panels) methods, as a function of [Fe/H]. Different Y values (0.25, 0.30, 0.35 and 0.40) 
and different input relative ages (0.62, 0.78, 0.94 and 1.09  for the horizontal method  and 0.62 and 
0.78 for the vertical one ) are considered. The horizontal and vertical parameters measured on 
He-enhanced isochrones are compared with the theoretical grids for the {\sl reference} isochrones 
shown in Figure~\ref{HeCNOEffect} to derive the age. \label{classicalHe}}
\end{figure}

\begin{figure}
\epsscale{1.00}
\plotone{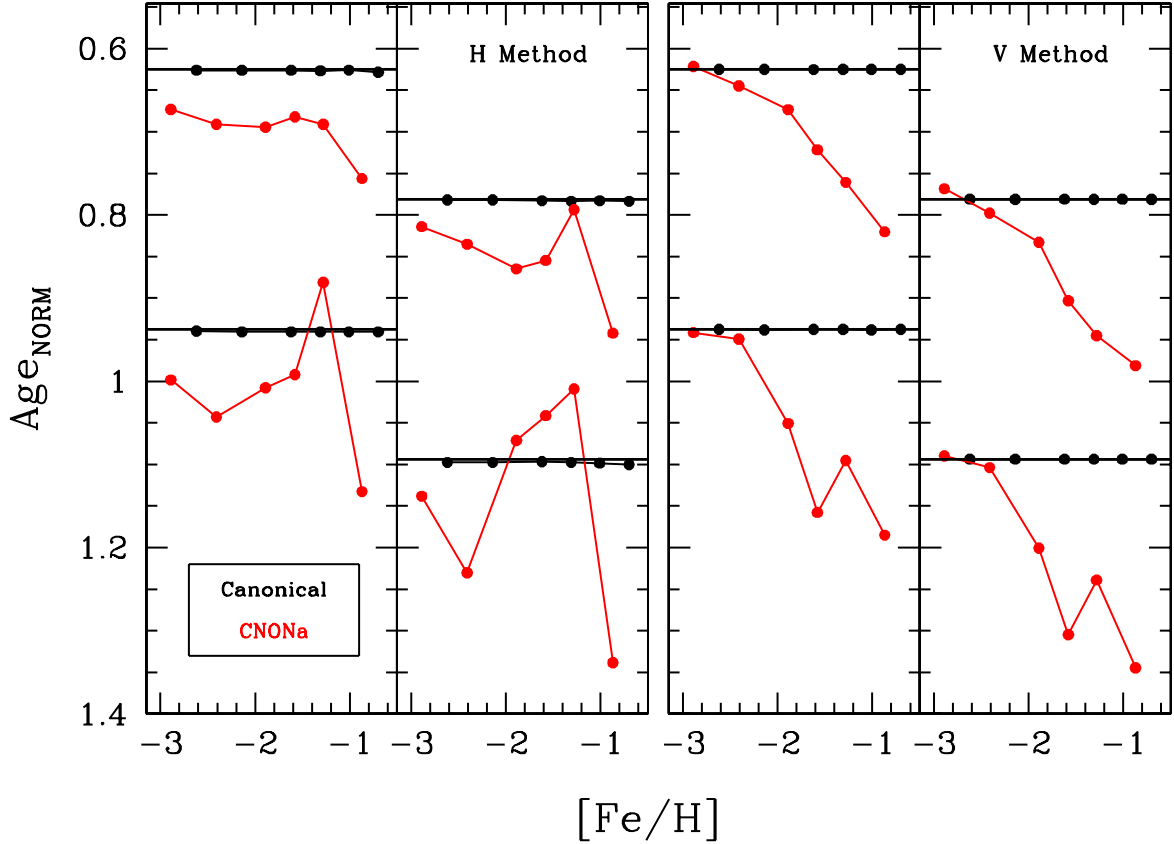}
\caption{Retrieved relative ages as a function of [Fe/H], measured using the horizontal and the vertical 
methods.  Two different CNONa mixtures and different input relative ages (0.62, 0.78, 0.94 and 1.09) have 
been used. The horizontal and vertical parameters measured on extreme CNONa isochrones, are compared with 
the theoretical grids for the {\sl reference} isochrones shown in Figure~\ref{HeCNOEffect} to derive the 
age. \label{classicalCNO}}
\end{figure}

\begin{figure}
\epsscale{1.00}
\plotone{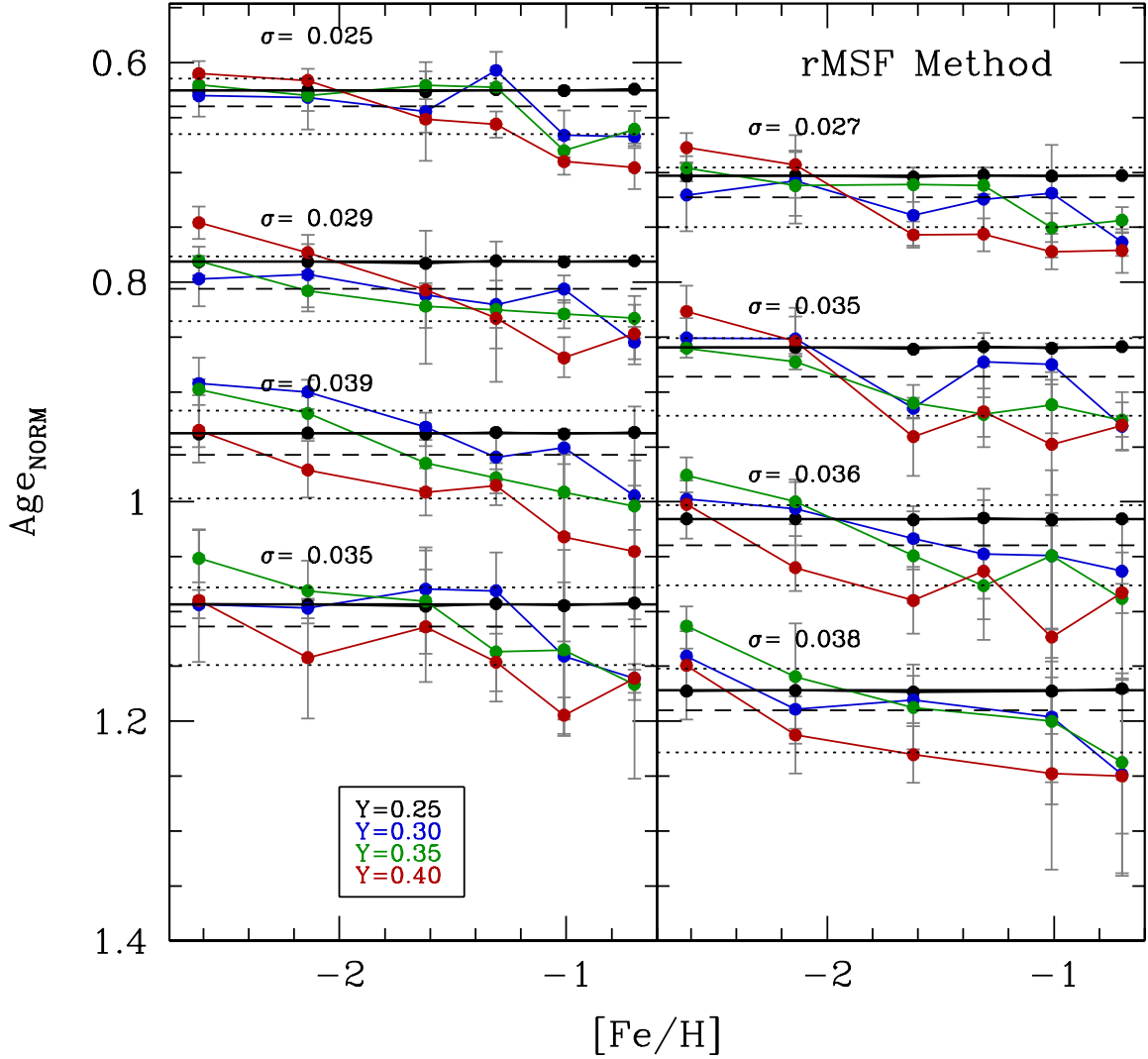}
\caption{Retrieved relative ages as a function of [Fe/H], obtained by using the rMSF method. Error bars 
represent the relative age uncertainty derived from the rMSF procedure. Different He values (0.25, 0.30, 
0.35 and 0.40) and various input relative ages (0.62, 0.70, 0.78, 0.86, 0.94, 1.02, 1.09 and 1.17) are 
considered. Measured parameters are compared with theoretical grid in Figure~\ref{HeCNOEffect} to derive 
the age. Dashed lines show the mean measured age and dotted lines show the 1-$\sigma$ interval. The value of 
$\sigma$ is listed for all input ages. \label{msFittingHe}}
\end{figure}

\begin{figure}
\epsscale{1.00}
\plotone{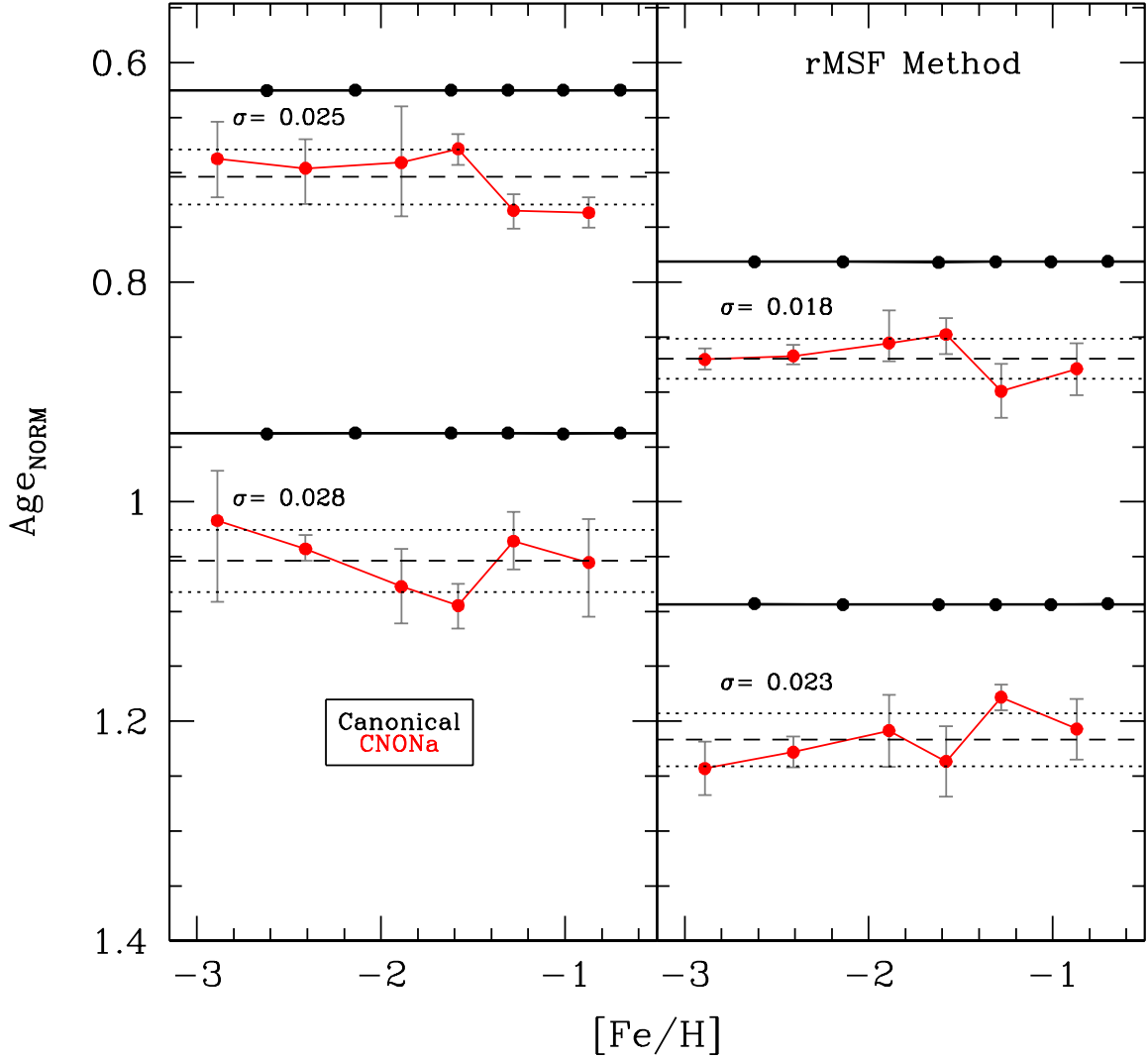}
\caption{Retrieved relative ages as a function of [Fe/H] based on the rMSF method. Two different CNONa 
mixtures and different input ages (0.62, 0.78, 0.94 and 1.09) have been used. Dashed lines show the mean 
measured age for the extreme CNONa mixture isochrones, and dotted lines show the 1-$\sigma$ interval. The 
value of $\sigma$ is listed for all input ages. \label{msFittingCNO}}
\end{figure}

\begin{figure}
\epsscale{1.00}
\plotone{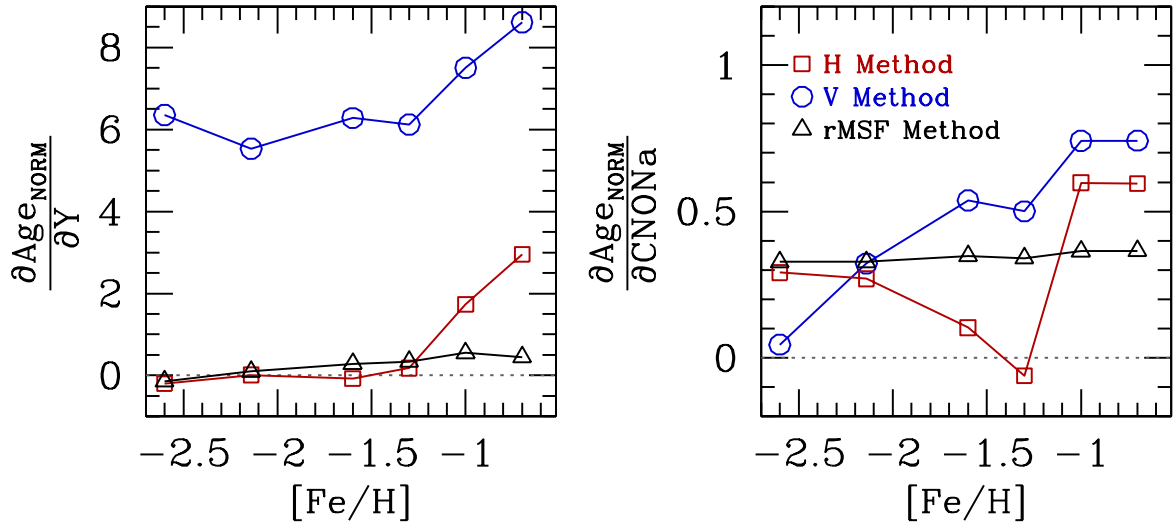}
\caption{Partial derivatives of the relative age with Y (left panel) and with CNONa (right panel) 
calculated for the horizontal (red), vertical (blue) and rMSF (black) methods. \label{deriva}}
\end{figure}

\end{document}